# Elongated focus optoacoustic microscopy with matched Bessel beam illumination and ultra-broadband axicon detection

**Zakiullah Ali**[1,2,3], **Christian Zakian**[1,2,3], **Vasilis Ntziachristos**[1,2,*]

[1] Chair of Biological Imaging, Central Institute for Translational Cancer Research (TranslaTUM), Technical University of Munich, Munich, Germany, 80333
[2] Institute of Biological and Medical Imaging, Helmholtz Zentrum München, Neuherberg, Germany, 85764
[3] Equal contribution
*Corresponding author: bioimaging.translatum@tum.de

# Abstract

Current optoacoustic microscopy configurations often have narrow focal ranges that limit their use for fast volumetric imaging applications. Here, the focal range of optoacoustic microscopes is extended by matching the elongated optical illumination profile of a Bessel beam with the pencil beam acoustic sensitivity profile of a broadband axicon detector. An inverted optoacoustic microscope was developed with interchangeable optical illumination and acoustic detection units to assess the imaging depths and resolutions retained with several combinations of illumination and detection profiles. Matching Bessel illumination with axicon detection extends the depth-of-focus 17-fold over traditional configurations. Imaging a tilted mouse ear with the matched Bessel-axicon configuration revealed vasculature over an imaging depth exceeding 4.2 mm with optical resolution, while affording a 6-fold increase in imaging volume over the same scanning duration compared to configurations employing standard Gaussian illumination, demonstrating this approach's promise for increasing applications for optoacoustic microscopy in preclinical research.

# Introduction

Optoacoustic (OA) microscopy is a label free hybrid approach that uses optical pulses to induce the emission of broadband sound waves in biological tissue, which are subsequently detected by an ultrasound transducer[1]. Common OA microscopy methods apply Gaussian beam illumination, which is diffraction limited and thus has a confined focal range (Rayleigh range). Spherically focused transducers are often employed for detection in OA microscopy because their spatial acoustic sensitivity profile complements the optical sensitivity profile of Gaussian illumination, which maximizes the overall optoacoustic sensitivity[2]. Despite the excellent sensitivity afforded by the combination of Gaussian beam illumination and spherically focused transducers, such OA microscopy configurations only provide high resolution imaging at the Gaussian beam focal point[3-5]. This narrow focus necessitates scanning the sample in the axial direction over the same focal plane to produce 3D images. Hence, current OA microscopy configurations are sub-optimal for volumetric imaging applications that are time sensitive or involve sample motion, such as full-body zebra fish imaging, measuring thick optically-cleared samples, or cell imaging and counting.

Unlike Gaussian beams, which have point foci, so-called ideal Bessel beams are not diffraction limited, and therefore have line-shaped foci. In practice, Bessel beams are approximated by directing collimated Gaussian illumination through a conical (axicon) lens, which yields a line shape optical sensitivity profile that enables imaging over extended depths without sacrificing lateral resolution[6,7]. As such, Bessel beam illumination has been investigated as a means of increasing imaging depth in several optical imaging modalities, including light sheet microscopy, optical coherence tomography, and second harmonic generation[8-13].

Attempts to increase imaging depth using with Bessel beam illumination have also been reported in optoacoustic microscopy. For example, an OA microscope using Bessel beam illumination was used to image red blood cells with 7 µm lateral resolution over a depth of 1 mm[14]. Similarly, two imaging studies of mouse ears demonstrated a 7-fold increase in imaging depth when imaging with Bessel beams in place of Gaussian beams. The lateral resolution/depths-of-focus of these systems were reported to be 1.6 µm/483 µm[15] and 300 nm/229 µm[16]. However, these microscopes employed either unfocussed or spherical transducers with acoustic sensitivity profiles mismatched to the Bessel beams' line foci, thereby limiting the optoacoustic sensitivity in the axial direction. This problem has been circumvented by employing optical axicon interferometric ultrasound detection paired with Bessel beam illumination to image zebra fish with a lateral resolution of 2.4 µm and a 635 µm depth-of-focus[17]. However, optical interferometric detection requires a thermally stabilized continuous wave light source and high speed balanced photodetector with high bandwidth, which increases the system's cost. In addition, optical ultrasound detection requires strict beam path alignment and environmental control, as the beam paths are sensitive to temperature fluctuations and micro-vibrations. Furthermore, trigger synchronization is required between the balanced photodetector driven by a secondary laser and the optoacoustic laser, thereby increasing the complexity of data acquisition. Finally, the interferometric signals measured by the photodetector require calibration and reconstruction to recover the original optoacoustic signal.

Our group recently demonstrated that a high-frequency ultra-wideband ultrasound transducer integrating an axicon lens to focus the incoming ultrasound waves along the axial direction offers extended acoustic depth-of-focus with retained high lateral resolution[18]. We postulated that the thin cylindrical axial sensitivity field (i.e. pencil beam sensitivity) of the axicon transducer would complement the optical profile of a Bessel beam, thus enabling a simple low cost approach to sense acoustic waves with improved sensitivity along the axial direction, potentially maximizing available imaging depths in OA microscopy and enabling fast volumetric imaging.

We therefore set out to establish whether the depth-of-focus in OA microscopy could indeed be maximized by matching the ultrasound detector sensitivity to Bessel beam illumination using an ultra-broadband axicon transducer, while also evaluating the effects of mismatches in illumination and detection geometry. Accordingly, we devised a custom OA microscope that allows modular exchange of illumination and detection units while retaining the same imaging field-of-view. We perform a robust and direct comparison of imaging depths and lateral resolutions achieved by four combinations of illumination (Gaussian or Bessel) and detection (spherical or axicon) geometries. In a controlled experiment, we demonstrate that our proposed configuration, with matched Bessel illumination and axicon detection, extends the OA microscopy depth-of-focus by up to 17 times compared to configurations that combined Gaussian illumination and spherically focused transducers. We highlight the expanded depth-of-focus, long working distance, and high-resolution of our optoacoustic microscope by imaging the vasculature of a tilted mouse ear over depths exceeding 4.2 mm. This extended depth-of-focus increased the imaging volume 6-fold over the same scanning duration compared to the Gaussian–spherical combination.

## Results

### Broadband spherical and axicon optoacoustic transducer characteristics

Two broadband single element side looking optoacoustic transducers of similar dimensions were manufactured and are depicted in Fig. 1(a & b). Both sensors housed a planar lithium niobate ($LiNbO_3$) active element. The acoustic beam is focused by mounting either a spherical (HFM36, Sonaxis, France) or axicon (HFM37, Sonaxis, France) lens onto the active element. Both transducers also feature a 2 mm central aperture suitable for central illumination. The spherical transducers has an 8 mm-diameter active element (central frequency = 60.5 MHz) and an acoustic lens with a curvature of 6 mm. The axicon transducer has a 9.6 mm-diameter active element (central frequency 61 MHz) and an acoustic lens with an apex angle of 114.4° (For further transducer information please refer to[18]). To assess the optoacoustic response for all four possible combinations of Gaussian and Bessel illumination with spherical and axicon detection, we designed an inverted optoacoustic microscope with interchangeable illumination optics and broadband transducers (Fig. 1c).

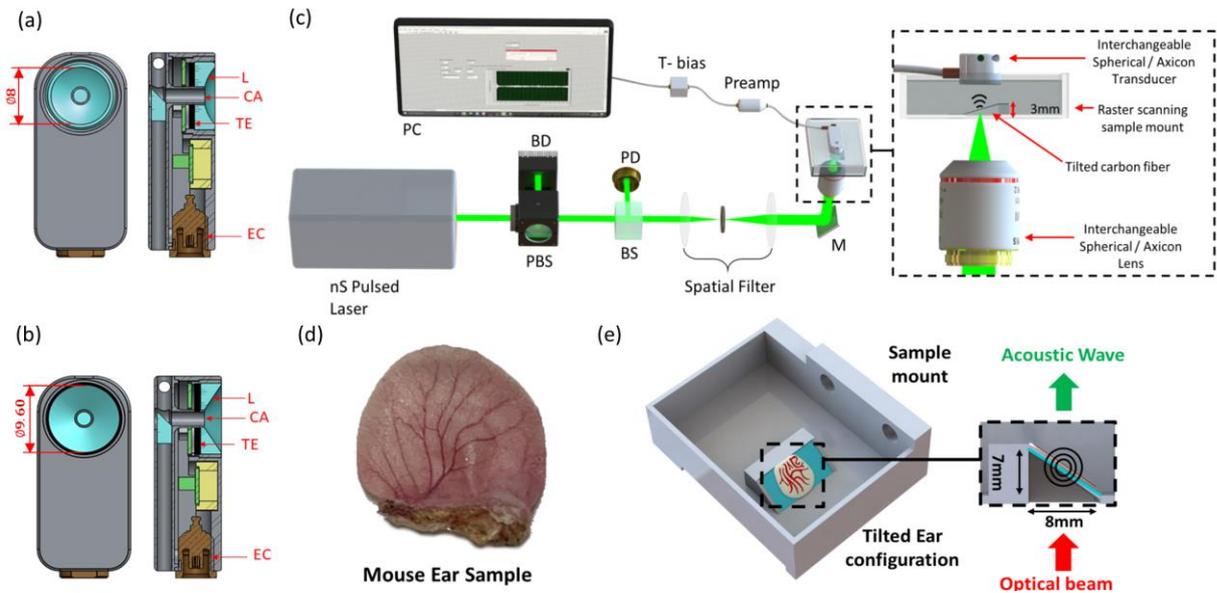

**Fig. 1:** Optoacoustic transducers and optoacoustic microscopy experimental setups for imaging a carbon fiber and ex-vivo mouse ear. Cross sectional and enface views of (a) HFM36 spherical and (b) HFM37 axicon transducers with central apertures. TE: Transducer element, EC: Electrical Connector, L: Acoustic Lens, CA: Central Aperture Window. (c) Depicts the optoacoustic microscope setup and tilted carbon fiber imaging configuration. PBS: Polarizing beam splitter, BS: beam splitter, BD: Beam dump, PD: Photodiode, M: Mirror. (d) An excised mouse ear sample sealed with a hot blade. (e) Illustration of the ex-vivo experimental setup wherein the mouse ear is placed on a tilted transparent optical window within a 3D printed tank. Fig. 1a-b reproduced from ref. 18 under CC BY 4.0.

## Gaussian and Bessel beam profile characteristics

The transverse illumination profile in optoacoustic microscopy determines the lateral resolution of the imaging system. Commonly, Gaussian beams are produced with tight foci to increase the resolution at the expense of light divergence away from the focal point. In contrast, Bessel beams are produced to create a transverse illumination profile concentrated in a central lobe, which can propagate long distances without diverging. To create the Gaussian or Bessel illumination patterns, we focused a collimated laser beam through either a spherical or an axicon lens, respectively. We performed optical beam profile measurements to calculate the beams' transverse intensity distributions along different points in the propagation direction. Fig. 2(a) and (b) show the resulting FWHM illumination beam diameters of approximately 5 µm for both the Gaussian beam and the central lobe of the Bessel beam. As expected, the Gaussian beam displays light confinement within the 5 µm-spot, whereas the Bessel beam shows a transversal pattern with lower amplitude concentric rings around the 5 µm central lobe. These concentric rings are an inherent property of Bessel beams, with each ring (including the central lobe) containing the same of energy. Hence the concentric rings generate secondary unwanted optoacoustic signals, which interfere with the signal generated from the central lobe and thereby degrade imaging lateral resolution. The Gruneisen relaxation effect[14] and blind deconvolution[15,16] have been exploited in optoacoustic microscopy to suppress Bessel beam side lobes and their associated artefacts. Here we also apply the blind deconvolution algorithm, as this method offers a computationally efficient and cost effective solution (see Fig 5). The beam profile and relative peak irradiance were measured along the optical axis for both illumination patterns. Fig. 2(c) shows the FWHM and relative peak irradiance of the Gaussian beam at 50 µm optical plane spacing intervals acquired along the optical axis over an axial distance of ±300 µm around the focal point. The peak irradiance is normalized to the maximum irradiance measured at focus. For the Gaussian illumination, we confirmed a minimum beam diameter of approximately 5 µm at focus and calculated a depth-of-focus of 108 µm. Fig. 2(d) shows the FWHM and relative peak irradiance of the Bessel beam central lobe at 1 mm optical plane spacing intervals acquired along the optical axis over an axial distance of 12 mm moving away from the axicon lens. The arbitrary origin during the course of the Bessel beam measurement was set to approximately 6 mm from the surface of the axicon lens. The Bessel beam diameter decreases with distance, reaching an average value of 5 μm over an 8 mm-range. The relative irradiance for the Bessel beam's central lobe increases with distance, reaching a plateau where the Bessel beam is fully formed at relative position of 8 mm.

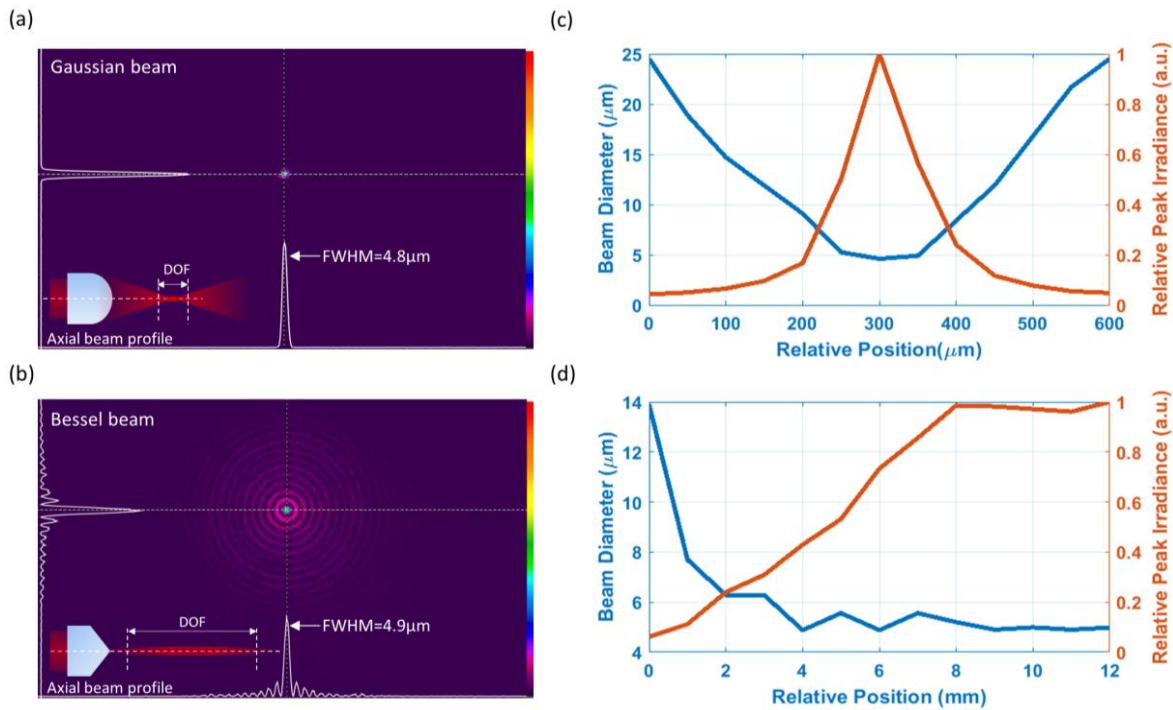

**Fig. 2:** Illumination beam profiles for optoacoustic microscopy. (a) Gaussian and (b) Bessel beam profiles measured at the lateral plane passing through the point of maximum irradiance. Relative peak irradiances and diameters of (c) a Gaussian beam measured at 50 μm optical planes through the optical focus. (d) A Bessel beam central lobe at 1 mm optical planes along the optical axis moving away from the axicon lens. Note: The arbitrary origin for the Bessel beam case is set to be 6 mm away from the surface of the axicon lens.

## Matched Bessel beam–axicon detection extends imaging depth-of-focus by an order of magnitude over standard Gaussian beam-spherical detection

To study the effect of matching illumination and detection profiles on imaging performance, we developed a customized OA microscope in transmission mode with exchangeable illumination and detection units, which enables imaging of a test target with the same field-of-view. We studied four possible combinations of Gaussian and Bessel illumination and spherical and axicon transducer detectors (Fig. 3a). Fig. 3b shows the intensity normalized maximum intensity projection (MIP) images of a tilted 7 μm-carbon fiber for all four illumination-detection configurations. Both, spherical and axicon detection, when combined with Gaussian illumination (Fig 3b(i-ii)), resolved the carbon fiber at the illumination focal point with a FWHM of 7 μm, while detecting the fiber over a total axial distance (depth-of-focus) of approximately $z = 225$ μm. The Bessel illumination combined with spherical detection (Fig 3b(iii)) resolved the carbon fiber at the point of highest optoacoustic intensity with a FWHM of 7 μm, while detecting the tilted fiber over the full field-of-view of 6 mm × 0.128 mm. However, the normalized optoacoustic sensitivity for the Bessel-spherical combination decreased significantly between $y = 0$ mm to $y = 1$ mm and between $y = 4$ mm to $y = 6$ mm, after which the carbon fiber is barely distinguishable. The normalized intensity for the Bessel-spherical combination was approximately 13% of the normalized maximum value at $y = 0.65$ mm and $y = 4.4$ mm, while the corresponding imaging depth in that range was $z = 1.98$ mm. This drastic decrease in intensity away from the center of the image is a function of the spherical shape of the detector, which concentrates sensitivity at the focal point. Finally, the combination of Bessel beam illumination with the axicon transducer (Fig. 3b(iv)) reveals the tilted carbon fiber over the complete depth of 3 mm with improved signal-to-noise ratio (SNR). Combining Bessel illumination with axicon detection results in retaining optoacoustic intensity above 30% of the normalized maximum intensity at y=0 and above 23% of the normalized maximum intensity at y = 6 mm. Notably, the FWHM was ~7 μm over almost the entire field-of-view (see supplementary Fig 1).

To visualize the depth-of-focus and corresponding lateral resolutions for the four imaging configurations, we generated false-color depth maps from the MIP images in Fig. 3b, bounded by the depth-of-focus and lateral resolution (Fig 3c). We computed the depth-of-focus by extracting the axial range at which the normalized MIP is above 50% and we calculate the lateral resolution as the FWHM across the carbon fiber for each imaging depth within the depth-of-focus. Irrespective of the transducer used, either spherical (Fig. 3c(i)) or axicon (Fig. 3c(ii)), the Gaussian illumination resulted in a depth-of-focus of 75 μm. The Bessel illumination used in combination with the spherical transducer (Fig. 3c(iii)) extends the depth-of-focus by 9-fold to 662 μm. As expected, the Bessel illumination combined with the axicon detection (Fig. 3c(iv)) maximizes the depth-of-focus to 1275 μm. Hence, the matched Bessel illumination and axicon detection results in a doubling of the depth-of-focus when compared to Bessel illumination with spherical detection, and a 17-fold increase in the depth-of-focus when compared to the Gaussian illumination combined with either of the transducers.

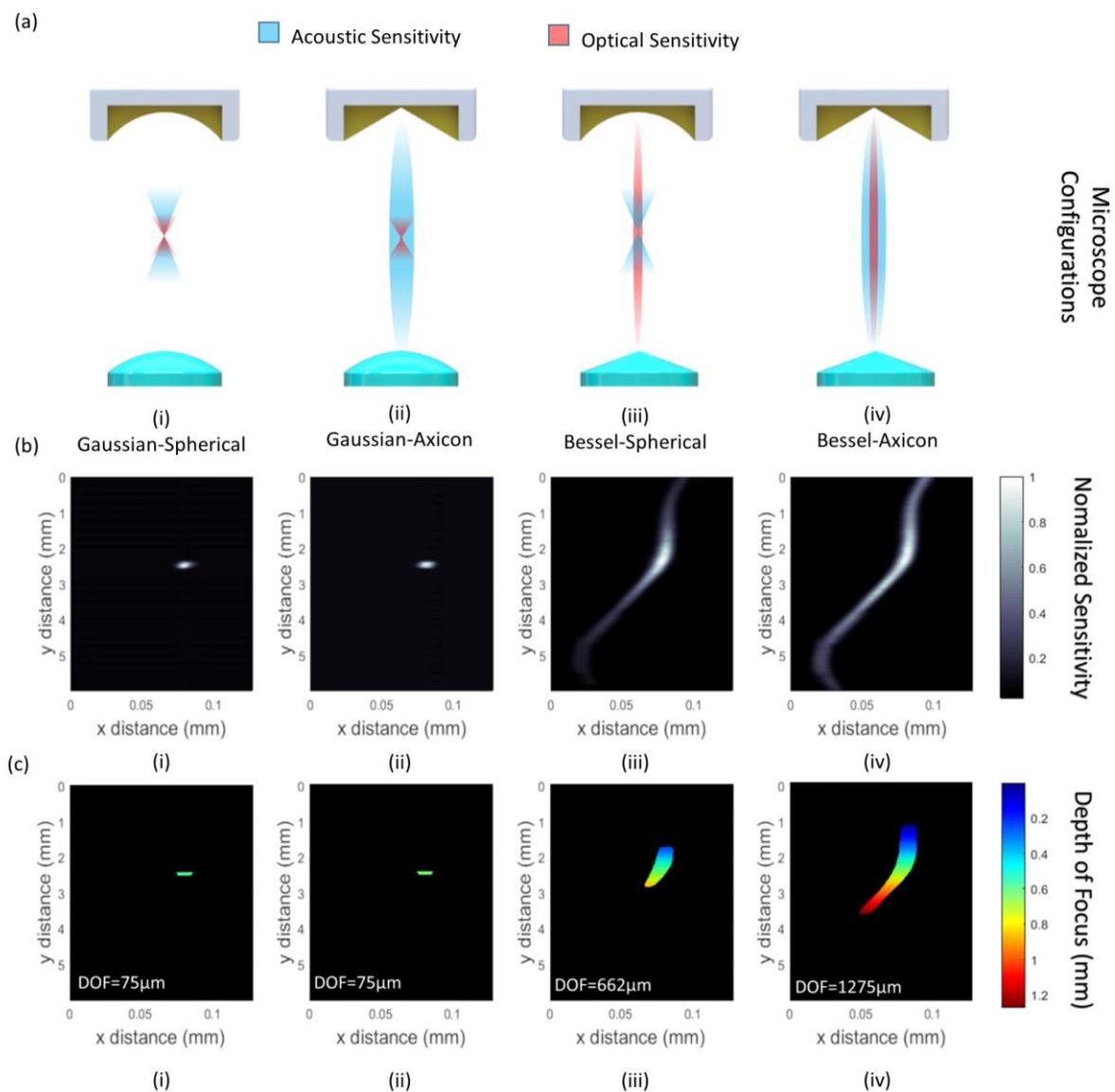

**Fig. 3:** Optoacoustic microscopy of a tilted 7 μm-carbon fiber using different acoustic and illumination configurations. (a) Illustrations of the four optoacoustic microscopy configurations: (i) Gaussian beam – spherical transducer, (ii) Gaussian beam – axicon transducer, (iii) Bessel beam –spherical transducer, (iv) Bessel beam – axicon transducer. (b) Depictions of the normalized maximum intensity optoacoustic images. (c) Corresponding false-color depth maps displayed within the depth-of-focus and lateral resolution for each imaging configuration.

# Matched Bessel beam–axicon detection extends imaging depth over several millimeters while retaining optical resolution in biological tissue

To evaluate the performance of our proposed OA microscope configuration in biological samples with irregular surfaces, we imaged a freshly excised mouse ear placed at a tilt (Fig 1d-e). Fig. 4a-c illustrate the 3D projection and 2D top- and side-view maximum intensity projections attained by imaging the titled mouse ear with the combination of Gaussian illumination and spherical detection. The vasculature is visible over a depth of about 700 µm and a cubic volume of 28 mm$^3$ (8 mm x 5 mm x 0.7 mm). Similar results were obtained with the combination of Gaussian illumination and axicon detection (Fig. 4d-f), with vessels visible over a depth of 700 µm and a cubic volume of 28 mm$^3$ (8 mm x 5 mm x 0.7 mm). In contrast, Bessel illumination and spherical detection enabled the visualization of vasculature over a depth of 2.8 mm and a cubic volume of 112 mm$^3$ (8 mm x 5 mm x 2.8 mm; Fig. 4g-i). Finally, Bessel illumination matched with the axicon detection reveals the entirety of the mouse ear vasculature over a depth of 4.2 mm and a cubic volume of 168 mm$^3$ (8 mm x 5 mm x 4.2 mm; Fig. 4j-l). The retention of high resolution over the full depth of the ear allows visualization of small vessels in the upper region of the scan that are otherwise not revealed by the other configurations. Additionally, the Bessel-axicon combination visualizes 6 times the volume in the same period compared to the Gaussian illumination combinations and 1.5 time the volume compared to the Bessel-spherical combination.

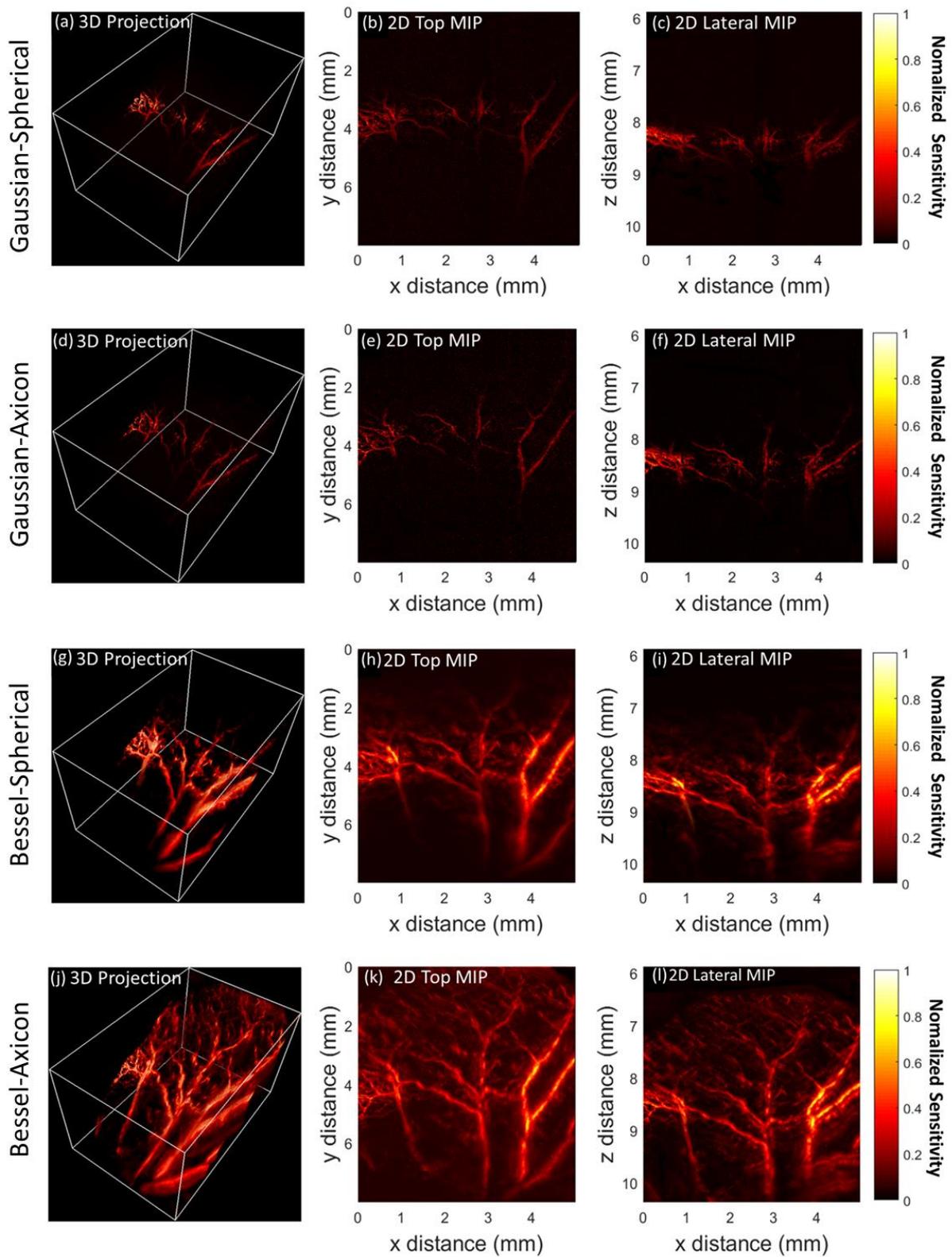

**Fig. 4:** Optoacoustic microscopy of a tilted mouse ear imaged with all four combinations of Gaussian or Bessel illumination and spherical or axicon transducers for detection. The sample was scanned using the same experimental configurations depicted in Fig. 3a(i-iv). 3D projections, 2D (top), and lateral (MIPs) of the mouse ear scanned with the following illumination/transducer combinations: (a-c) Gaussian–spherical, (d-f) Gaussian–axicon, (g-i) Bessel–spherical, and (j-l) Bessel–axicon.

# Deconvolution improves lateral resolution of optoacoustic images acquired with matched Bessel beam–axicon detection

To reduce image artifacts arising from the secondary side lobes of the Bessel illumination beam, MATLABs blind deconvolution algorithm was employed to enhance resolution and reduce background noise (Fig. 5). The raw MIP of the mouse ear acquired with the combination of Bessel illumination and axicon detection is shown in Fig. 5a. The corresponding deconvolved projection presented in Fig. 5b was attainted by deconvolving the original image (Fig. 5a) with the Bessel beam profile measurement (Fig. 2b). The dotted white lines in both Fig. 5a and 5b indicate the location of the line profile measurement for the raw and deconvolved data, shown in Fig. 5c. The line profiles show a 10% reduction in the background signal, resulting in an SNR improvement of 2.1 dB, and improving the vessel width definition from 132 µm to 75 µm after deconvolution.

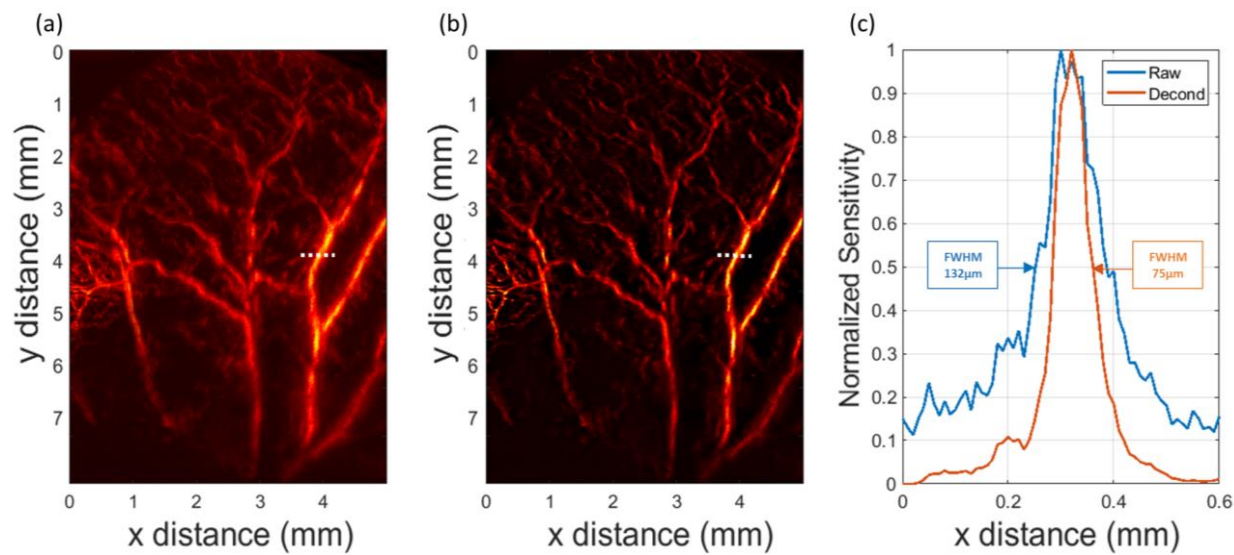

**Fig. 5:** Deconvolution of an optoacoustic microscopy MIP image of the vasculature of a mouse ear from the combination of Bessel illumination and axicon detection. (a) The raw and (b) deconvolved MIP images. (c) A comparison of FWHM measurements between the raw and deconvolved images, showing an improvement in image resolution and reduction in background noise.

## Discussion

Current OA microscopy configurations are sub-optimal for volumetric imaging or time sensitive applications because they employ Gaussian illumination restricting imaging depth within the Gaussian beam's Rayleigh range, which can extend up to a few tens of microns to ensure high lateral resolution at the focal plane. We present a new OA microscope configuration that allows detection over ultra-long depths-of-focus while retaining high lateral resolution over several millimeters. We show that the combination of Bessel illumination and optoacoustic axicon detection enables one order of magnitude higher imaging depth-of-focus compared to Gaussian illumination-based optoacoustic microscopy and doubles the imaging depth-of-focus when compared to an un-matched combination of Bessel illumination and focused detection with a spherical transducer. This expanded depth-of-focus increased the cubic volume acquired 6-fold when compared to Gaussian illumination optoacoustic microscopy over the same scanning duration and without adjusting the imaging plane to produce image stacks. Such capability could significantly accelerate image acquisition for time-sensitive measurements.

The combination of Bessel illumination with matched axicon detection maximizes the depth-of-focus of an optoacoustic microscope without sacrificing lateral resolution. Previous attempts to increase depths-of-focus in optoacoustic microscopy by using Bessel illumination[14-16] have been hampered by the use unfocussed or spherical transducers for detections, which limit SNR and sensitivity along the axial direction. Alternatively, optoacoustic signal detection via optical interferometry with Bessel beam illumination falls victim to stringent beam path alignment and environment control as the difference between optical paths are sensitive to vibrations and ambient temperature. Furthermore, additional light sources and high-speed photodetectors are required increasing system complexity and cost[17]. Here we demonstrate a simple low cost approach to increase imaging depth-of-focus, and thereby the accessible imaging volume, by matching the Bessel beams optical sensitivity profile with our axicon transducer's thin cylindrical acoustic sensitivity response, affording a 17-fold larger depth-of-focus compared to combinations of Gaussian illumination and either spherical or axicon detection (Fig. 3c). When imaging an excised tilted mouse ear, the Bessel beam-axicon transducer configuration allowed images of the entire vasculature network over a depth of 4.2 mm and cubic volume of 168 mm$^3$ (Fig 4j-l). In contrast, combining Bessel beam with a spherical transducer reduced the measured depth to 2.8 mm and cubic volume of the ear vasculature to 112 mm$^3$ (Fig. 4g-i), while using Gaussian illumination with either detector limited the overall imaging depth to 700 μm and the cubic volume to 28 mm$^3$ of the ear vasculature (Fig. 4a-f). Hence, the Bessel-axicon combination imaged 6-fold the cubic volume of the ear vasculature over the same scanning duration when compared to the combinations that employed Gaussian illumination.

While Bessel beam illumination enables extension of the focal depth without comprising the inner lobe spot size, the secondary lobes surrounding the inner lobe generate undesired optoacoustic signals deteriorating the overall lateral resolution of the optoacoustic image. The lateral resolution in Bessel beam-axicon transducer matched OA microscope configuration can be improved by deconvolving the OA image with the point spread function of the Bessel illumination. We found that the side lobes of the Bessel beam (Fig. 2b) indeed result in secondary optoacoustic signals, which interfere with the signal generated by the inner lobe, degrading the lateral resolution and contrast. Recent endeavors to suppress side lobe artifacts when using Bessel illumination include exploitation of the nonlinear Gruneisen relaxation effect, blind deconvolution, or interferometric detection[14-17]. However, the nonlinear Gruneisen process increases system cost and interferometric detection increases experimental complexity. Hence, we follow a similar approach to Jiang, B., et al[15], and performed post-processing suppression of the side lobes via blind deconvolution, which is computational efficient and cost effective. We deconvolved the acquired optoacoustic images with the optical beam profile measurement of the Bessel beam (Fig. 2b) and improve the resolution measured across a blood vessel (Fig 5) by a factor of 1.7 and also suppress background signal by an SNR improvement of 2.1 dB.

Using axicon acoustic detection for microscopy offers several advantages. For instance, even in combination with Gaussian illumination, axicon detection enables elongated acoustic sensitivity fields where the illumination could be focused by means of a spatial light modulator (SLM) at different positions along the axial direction. As a result, high resolution volumetric imaging could be obtained without mechanically moving the sample or imaging unit. Matching axicon ultrasound detection with Bessel illumination provides three main advantages. First, the configuration is robust to axial miss-alignment of the illumination, detector, and sample. Given the extended overlapping range of the matched illumination and detection sensitivity, the relative axial alignment can be coarse and does not affect image lateral resolution. Second, the expanded depth-of-focus enables high resolution imaging of irregular surfaces, where structures would otherwise be only partially imaged by Gaussian illumination and spherically focused detectors. Hence, Bessel illumination employed with axicon transducers are ideal for volumetric imaging in life-sciences applications such as OA imaging of full-body zebra fish[5,19], OA endomicroscopy[20,21] and OA flow cytometry[22,23]. Finally, fast OA microscopy in volumetric and uneven surfaces is possible as axial adjustment is not required and fast dynamic processes occurring in and out of plane could be imaged[24,25].

We have previously shown that the spherical transducer used in this study is 2.8 times more sensitive than its axicon counterpart, but that the axicon transducer offers 4.2 times the depth-of-focus[18]. Moreover, Bessel beams possess a cylindrically symmetric intensity profile that does not change in free space, with a central lobe that is remarkably resistant to diffractive spreading compared to that of Gaussian beams of similar beam diameter[7]. Hence, our proposed OA microscopy configuration distributes both the excitation energy density and the acoustic detection sensitivity along an extended axial direction. The elongated sensitivity profile results in weaker optoacoustic signal generation and reduced detection sensitivity compared to Gaussian illumination and spherical ultrasound detection. Nevertheless, the trade-off is most severe at the illumination and detection focal points and rapidly drops outside the focal plane, limiting the possibility to resolve out of plane structures without needing to adjust the foci to the plane of interest. Our configuration can uniquely resolve microscopic optoacoustic sources distributed over large working distances.

Optoacoustic microscopy imaging depth could be further extended by increasing the size of the axicon detector or removing the central transducer aperture, while the lateral resolution could be increased by reducing the size of the Bessel illumination central lobe by adopting an axicon lens with a larger physical angle. However, reducing Bessel beam central lobe diameter would also reduce the Bessel beam optical depth-of-focus[6,7]. Future studies will focus on exploiting the central aperture of the axicon detector by designing an axicon probe to be housed within the transducer and thereby enable reflection mode Bessel beam optoacoustic microscopy with an axicon transducer for in-vivo applications. In addition, further investigations will follow to develop a thin optical pencil beam without secondary side lobes to avoid the post processing deconvolution process. Moreover, a built-in pre-amplifier would increase the sensitivity of the axicon transducer and improve SNR.

In conclusion, by carefully matching Bessel illumination to a broadband axicon optoacoustic transducer, we achieve unprecedented imaging depths-of-focus for optoacoustic microscopy while retaining high lateral resolution. Although, lateral resolution remains a function of the illumination spot size, we demonstrate that the depth-of-focus is limited by the illumination when using focused Gaussian beams, but otherwise limited by the ultrasound detection sensitivity field when Bessel beams are employed. Hence, our proposed configuration significantly reduces the compromise between imaging depth and resolution, achieving lateral resolutions of 7 µm with imaging depths greater than 4.2 mm. The expanded depth-of-focus with high lateral resolution opens exciting opportunities for future biological and clinical applications, including deep microscopic molecular image of full samples and tissues with irregular surfaces.

## Methods

### Inverted optoacoustic microscope configuration

Our inverted microscope illustrated in Fig. (1c) consists of a light source with a 532 nm-wavelength and a 1 ns-pulse width (Wedge HB532, Bright Solutions, Italy). The energy per pulse was regulated by means of a polarizing beam splitter and a beam dump. A 90:10 beam splitter was employed to divert 10% of the light towards a photodetector for triggering. The remaining 90% of light was spatially filtered to clean and expand the beam prior to optoacoustic excitation. Gaussian illumination with a spot size of approximately 5 µm was achieved by employing a 0.25 NA 10X plan achromat (RMS10X, Olympus, Japan). Alternatively, axicon illumination with approximately 5 µm FWHM was realized by interchanging the objective lens with an axicon lens (AX255 UVFS, Thorlabs, USA).

The Gaussian and Bessel illumination profiles were characterized by measuring the beam profiles along the optical axis moving away from the illumination lens surface using a beam profiler (SP620U, Ophir, USA) with a secondary 10X plan achromat lens (RMS10X, Olympus, Japan) and an 80 mm-optical spacer. Fig. 2(a & b) illustrate the Gaussian and Bessel beam profiles at maximum irradiance along the optical axis, respectively. The FWHM and relative peak irradiances were measured along the optical axis to assess the illumination spatial confinement for each configuration. Fig. (2c) shows the Gaussian beam confinement measured at 50 µm optical plane intervals over an axial distance of ±300 µm around the focal point. Fig. (2d) illustrates the Bessel beam confinement measured at 1 mm optical plane intervals over an axial distance of 12 mm moving away from the axicon lens surface. For the Bessel beam experiment, the arbitrary origin was set to approximately 6 mm from the surface of the axicon lens.

## Tilted carbon fiber experimental arrangement and parameters

For OA imaging, each acoustic transducer was mounted to face the illumination source and aligned with the optical beam using a manual XYZ translation stage (Fig. 1c). All four possible illumination and acoustic detection combinations (Gaussian-spherical, Gaussian-axicon, Bessel-spherical, and Bessel-axicon) were evaluated by imaging a 7 µm-wide carbon fiber immersed in deionized water and fixed on a tilt with a height of 3 mm and an inclination angle of 23.2º inside a custom 3D printed tank with a planar optical window for transmission mode-illumination optoacoustic microscopy (Fig. 1c). The tank was attached to a motorized XYZ scanner for precise adjustment of the carbon fiber along the illumination-detection axis with a linear stage for the X scanning direction (M-511, Physik Instrumente, Germany) and two compact precision linear stages for the Y and Z scanning directions (MTS50-Z8, Thorlabs, USA). Upon alignment of the optical beam, transducer, and carbon fiber, a raster scan was performed for each illumination and detection combinations (Fig. 3a (i-iv); scan speed: 2 mm/s, scan resolution: 500 nm × 800 nm, field-of-view: 6 mm × 0.128 mm). The cross section of the Bessel beams consists of concentric rings, each containing the exact same amount of energy; therefore, only a small fraction of the total energy in Bessel beams remains within the central lobe. In our experiments, pulse illumination energies of 10 nJ and 2 µJ were used for the Gaussian and Bessel beams, respectively. The temporal optoacoustic signals were externally pre-amplified with a 30 dB amplifier (Sonaxis SA, France) and connected to a T-bias (ZFBT-4R2GW+, Mini-Circuits, USA) prior to data acquisition. Optoacoustic signals were acquired using a DAQ card (ATS9373, AlazarTech, Canada) synchronized to the external trigger port via the photodetector signal. Data acquisition was controlled by LabVIEW (National Instruments, USA) and data processing to generate 2D projections was performed in MATLAB (MathWorks, USA). 3D volumetric projections were generated with Amira-Avizo (Thermo fisher scientific, USA).

## Bessel beam side lobe suppression by point spread function deconvolution

Importantly, Bessel beam optical profiles also exhibit secondary concentric rings surrounding an inner lobe due to their non-diffracting nature, which generates unwanted secondary optoacoustic signals interfering with the signal generated from the central lobe, thereby degrading lateral image resolution. To suppress image artifacts and background signal arising from the secondary lobes of the Bessel beam, MATLAB's built in blind-deconvolution algorithm (deconvblind) was applied to the image projection acquired for the Bessel-axicon and spherical configurations. To improve the deconvolution, the Bessel beam profile (see Fig. 2b) was used as input for initial estimation of the point spread function(PSF) . The blind deconvolution technique has been previously validated in the field of optoacoustic microscopy to suppress side lobe artifacts [15,16]

## Tilted mouse ear experimental arrangement and parameters

Measurements of an ex-vivo mouse ear excised with a hot blade to seal the blood within (Fig. 1d) were performed in transmission mode with the sample placed on a tilted cover slit installed inside a customized 3D printed water tank at a height of 7 mm and an inclination angle of 41.2º as depicted in Fig. 1e. All four optical illumination and acoustic detection combinations were used to image the tilted mouse ear (scan speed: 2 mm/s, scan resolution: 1 µm × 10 µm, field-of-view of 8 mm × 5 mm). The pulsed energies were set to approximately 80 nJ and 8 µJ for Gaussian and Bessel beam illuminations, respectively.

# References

1. Ntziachristos, V. Going deeper than microscopy: the optical imaging frontier in biology. Nat Methods 7, 603-614, doi:10.1038/nmeth.1483 (2010).

2. Omar, M., Aguirre, J. & Ntziachristos, V. Optoacoustic mesoscopy for biomedicine. Nature Biomedical Engineering 3, 354-370, doi:10.1038/s41551-019-0377-4 (2019).

3. Kellnberger, S. et al. Optoacoustic microscopy at multiple discrete frequencies. Light: Science & Applications 7, 109, doi:10.1038/s41377-018-0101-2 (2018).

4. Pleitez, M. A. et al. Label-free metabolic imaging by mid-infrared optoacoustic microscopy in living cells. Nature Biotechnology 38, 293-296, doi:10.1038/s41587-019-0359-9 (2020).

5. Soliman, D., Tserevelakis, G. J., Omar, M. & Ntziachristos, V. Combining microscopy with mesoscopy using optical and optoacoustic label-free modes. Scientific Reports 5, 12902, doi:10.1038/srep12902 (2015).

6. Born, M. & Wolf, E. Principles of Optics: Electromagnetic Theory of Propagation, Interference and Diffraction of Light. 7 edn, (Cambridge University Press, 1999).

7. Durnin, J., Miceli, J. J. & Eberly, J. H. Comparison of Bessel and Gaussian beams. Optics Letters 13, 79-80, doi:10.1364/OL.13.000079 (1988).

8. Collier, B. B., Awasthi, S., Lieu, D. K. & Chan, J. W. Non-Linear Optical Flow Cytometry Using a Scanned, Bessel Beam Light-Sheet. Scientific Reports 5, 10751, doi:10.1038/srep10751 (2015).

9. Vuillemin, N. et al. Efficient second-harmonic imaging of collagen in histological slides using Bessel beam excitation. Scientific Reports 6, 29863, doi:10.1038/srep29863 (2016).

10. Lu, R. et al. Video-rate volumetric functional imaging of the brain at synaptic resolution. Nature Neuroscience 20, 620-628, doi:10.1038/nn.4516 (2017).

11. Fahrbach, F. O., Simon, P. & Rohrbach, A. Microscopy with self-reconstructing beams. Nature Photonics 4, 780-785, doi:10.1038/nphoton.2010.204 (2010).

12. Lee, K.-S. & Rolland, J. P. Bessel beam spectral-domain high-resolution optical coherence tomography with micro-optic axicon providing extended focusing range. Optics Letters 33, 1696-1698, doi:10.1364/OL.33.001696 (2008).

13. Planchon, T. A. et al. Rapid three-dimensional isotropic imaging of living cells using Bessel beam plane illumination. Nat Methods 8, 417-423, doi:10.1038/nmeth.1586 (2011).

14. Shi, J., Wang, L., Noordam, C. & Wang, L. Bessel-beam Grueneisen relaxation photoacoustic microscopy with extended depth of field. Journal of Biomedical Optics 20, 116002 (2015).

15. Jiang, B., Yang, X. & Luo, Q. Reflection-mode Bessel-beam photoacoustic microscopy for in vivo imaging of cerebral capillaries. Opt. Express 24, 20167-20176, doi:10.1364/OE.24.020167 (2016).

16. Park, B. et al. Reflection-mode switchable subwavelength Bessel-beam and Gaussian-beam photoacoustic microscopy in vivo. Journal of Biophotonics 12, e201800215, doi:10.1002/jbio.201800215 (2019).

17. Hu, Y., Chen, Z., Xiang, L. & Xing, D. Extended depth-of-field all-optical photoacoustic microscopy with a dual non-diffracting Bessel beam. Optics Letters 44, 1634-1637, doi:10.1364/OL.44.001634 (2019).

18. Ali, Z., Zakian, C. & Ntziachristos, V. Ultra-broadband axicon transducer for optoacoustic endoscopy. Scientific Reports 11, 1654, doi:10.1038/s41598-021-81117-7 (2021).

19. Moore, M. J. et al. Simultaneous ultra-high frequency photoacoustic microscopy and photoacoustic radiometry of zebrafish larvae in vivo. Photoacoustics 12, 14-21, doi:https://doi.org/10.1016/j.pacs.2018.08.004 (2018).

20. Yang, J.-M. et al. Optical-resolution photoacoustic endomicroscopy in vivo. Biomedical Optics Express 6, 918-932, doi:10.1364/BOE.6.000918 (2015).

21. Xiong, K., Wang, W., Guo, T., Yuan, Z. & Yang, S. Shape-adapting panoramic photoacoustic endomicroscopy. Optics Letters 44, 2681-2684, doi:10.1364/OL.44.002681 (2019).

22. Nedosekin, D. A. et al. Photoacoustic flow cytometry for nanomaterial research. Photoacoustics 6, 16-25, doi:https://doi.org/10.1016/j.pacs.2017.03.002 (2017).

23. Cai, C. et al. Photoacoustic Flow Cytometry for Single Sickle Cell Detection In Vitro and In Vivo. Anal Cell Pathol (Amst) 2016, 2642361-2642361, doi:10.1155/2016/2642361 (2016).

24. Zhang, P. et al. High-resolution deep functional imaging of the whole mouse brain by photoacoustic computed tomography in vivo. Journal of Biophotonics 11, e201700024, doi:https://doi.org/10.1002/jbio.201700024 (2018).

25. Gottschalk, S. et al. Rapid volumetric optoacoustic imaging of neural dynamics across the mouse brain. Nature Biomedical Engineering 3, 392-401, doi:10.1038/s41551-019-0372-9 (2019).

## Acknowledgements

We thank Robert J. Wilson for discussions and editing the manuscript. The authors would like to acknowledge Mr. Guillaume Pierre (SONAXIS SA, France) for his support in the design and manufacturing of the transducers according to specifications. This project has received funding from the European Union's Horizon 2020 research and innovation programme under grant agreement No 732720 (ESOTRAC), as well as from the European Union's Horizon 2020 research and innovation programme under the Marie Sklodowska-Curie grant agreement No 721766 (FBI)

## Author contributions

All experiment realization and figure designs: Z.A.
Supervision and conception: C.Z.
Analysis, interpretation of data and drafting of manuscript: Z.A, C.Z, V.N
Obtaining funding, drafting of manuscript: V.N

## Conflict of Interest statement

The authors declare no conflicts of interest.

## Data availability

The datasets generated during and/or analysed during the current study are available from the corresponding author on reasonable request.

# Supplementary Information

## FWHM measurements of tilted suture phantom with Bessel beam illumination

Supplementary Figure. 1 depicts the FWHM measurements of the tilted 7 µm carbon fiber with the Bessel beam illumination used in combination with the spherical and axicon detectors. In general, an approximate FWHM of 7 µm is measured over a large field-of-view. The variation in FWHM measurements could be a result of the tilted fiber slightly moving in water during the scanning procedure.

The degraded lateral resolution measured between y = 5 mm to y = 6 mm of the maximum intensity plots of approximately 10 µm (Bessel – spherical combination) and 9 µm (Bessel – axicon combination) is attributed to a lower SNR as the source of acoustic signal is out of the depth-of-focus of the transducers and the generated source also experiences an extended propagation distance loss. Furthermore, there is a sharp bend in the carbon fiber in this region, which further degrades the lateral resolution.

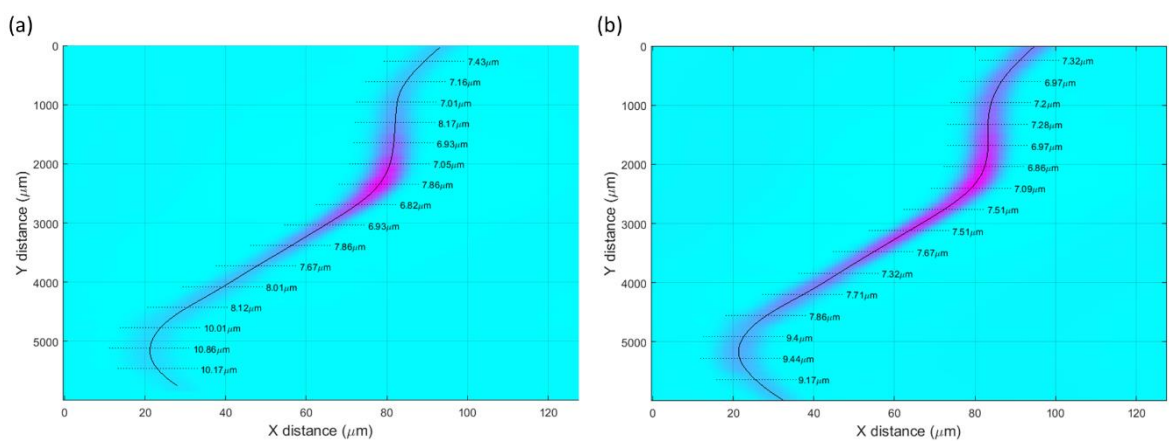

**Supplementary Figure 1.** FWHM measurements of the tilted 7 µm carbon fiber with Bessel illumination used in combination with the spherical and axicon detectors. (a) Bessel illumination – spherical detection configuration (b) Bessel illumination – axicon detection configuration.